\documentclass[aps,prev,twocolumn,preprintnumbers,floatfix,nofootinbib]{revtex4}
\usepackage{epstopdf}
\usepackage{mathrsfs}
\usepackage{amssymb}
\usepackage{verbatim}
\usepackage{epsf, epsfig, subfigure, graphicx, amsmath, amssymb}
\usepackage{color}

\newcommand{\GeV}{{\rm GeV}}

\begin{document}

\title{Two Component Higgs-Portal Dark Matter}
\author{Ligong Bian}
 \email{lgb@mail.nankai.edu.cn}
\affiliation{%
School of Physics, Nankai University, Tianjin 300071, P. R. China
}%

\author{Ran Ding}
 \email{dingran@mail.nankai.edu.cn}
\affiliation{%
School of Physics, Nankai University, Tianjin 300071, P. R. China
}%

\author{Bin Zhu}
 \email{zhubin@mail.nankai.edu.cn}
\affiliation{%
School of Physics, Nankai University, Tianjin 300071, P. R. China
}%

\begin{abstract}
In this paper, we construct two component dark matter model and revisit fine-tuning, unitarity and vacuum stability problem in this framework. Through Higgs-portal interactions, the additional
scalar and vector singlet field can interact with the SM particles. The parameter space of the model are severely constraint by observed relic density and direct detection experiments. We found that, unlike the SM, the fine-tuning problem is relaxed due to the modified Veltman condition. The vacuum stability problem is addressed, the additional contributions from two DM singlets to the $\beta$ function make the Higgs quartic coupling $\lambda(\mu)$ be positive up to Planck scale in some parameter space.
\end{abstract}
\maketitle
\section{Introduction}
The discovery of $125$ GeV Standard Model(SM)-like Higgs boson~\cite{Aad:2012tfa} indicates that the SM seems to be complete. However, there exist overwhelming evidences which convince us that the SM is at best an effective theory valid up to some energy scale. Among them, two of most intriguing evidences are the fine-tuning problem of the Higgs mass and the existence of non-baryonic dark matter (DM).
In the context of the SM, the conventional fine-tuning  problem could be expressed as follows
 \begin{eqnarray}
 (m^{0}_{h})^{2}=m^{2}_{h}+
 \frac{\Lambda^{2}}{(4\pi)^{2}v^{2}}VC_{SM}\;,
 \end{eqnarray}
 with
 \begin{eqnarray}
 VC_{SM}\equiv6m^{2}_{h}-24m^{2}_{t}+12m^{2}_{W}+6m^{2}_{Z}\;.
 \end{eqnarray}
In the above formula, $m_{h},m_{t},m_{W},m_{Z}$  are renormalized Higgs, top-quark,
 W and Z boson mass respectively, $v$ is the corresponding vacuum expectation value (VEV), $m^{0}_{h}$
 is the bare Higgs mass,
 and the cut off scale $\Lambda$ indicates where new physics enters into SM.
The contributions from other quarks and leptons are neglected safely compared with that of top quark.
In order to relax fine-tuning, we require
 \begin{eqnarray}\label{eq:FTSM}
 m^{2}_{h}\geq\frac{\Lambda^{2}}{(4\pi)^{2}v^{2}}VC_{SM}\;.
 \end{eqnarray}
Thus we obtain $\Lambda\sim 546$ GeV.
No new physics has been observed sofar at LHC which indicates that the cutoff might be larger
than several TeV. Then new fields are required to soften the fine-tuning problem~\cite{Grzadkowski:2009mj,Chakraborty:2012rb}.\footnote{The simplest solution to
the problem is to suppose $VC_{SM}=0$~\cite{Veltman:1980mj}, the conventional Veltman condition
is realized.
It is obvious that this condition could not be satisfied in the SM at low scale,
since the top-quark's contribution is larger than the others in this case,
giving rise to negative contribution to $VC_{SM}$.} In the paper of~\cite{Chakraborty:2012rb},
the field content of
the SM is extended by a scalar singlet and a vector fermion field, where
 the later one
only interacts with the $S$ through $\bar{F}FS$. While for absence of tadpole
contributions, the self-energy come from $\bar{F}FS$ and $H^{\dag}HS$
interaction are dropped. With the scalar singlet served as dark matter
candidate, it was found that the
fine-tuning of the Higgs mass is a strict
constraint for the Higgs-portal coupling, and the number of scalar singlets $N$,
which is set to be larger than $10$ through direct detection constraint, i.e., XENON100 experiment. This
$O(N)$ symmetry should be broken softly and make the lightest scalar $S$ to be DM candidate.
The novelty here is that the tadpole
contributions are non-negligible to respect gauge invariance of the Higgs two-point Green Function~\cite{Ma:1992bt,Fleischer:1980ub}.

The straightforward generalization of the above idea is introducing additional fermionic DM, which can interact with SM through Higgs-portal.
To improve the fine-tuning problem, the new fields entering into the Veltman condition should be bosonic fields which provide positive contributions. On the contrary, the fermionic
fields give negative contributions thus makes the problem worse.

In this work, we consider both the scalar $S$ and vector field $V$ as Dark matter candidate, both of them can
annihilate into SM particles and each other through Higgs-portal interaction. Total DM relic density and direct direction experiments impose the stringent constraints on parameter space of our model. We then use allowed parameter space to revise the fine-tuning, unitarity and the vacuum stability of Higgs boson.

The paper is organized as follows. In section II, we extend the conventional Higgs-portal model to include two component DM candidates. DM relic abundances calculation and experiment constraints are present in section III. The fine-tuning, unitarity
and the vacuum stability of Higgs boson are examined in section IV. Finally, the last section is devoted to conclusions.

\section{Two component dark matter}

To explain the existence of DM, the SM must be extended.
The minimal model is to add the Higgs-portal interactions, such as the singlet scalar
, fermionic or vector field as dark matter candidate
~\cite{Djouadi:2011aa,Djouadi:2012zc,Eboli:2000ze,Kanemura:2010sh,Masina:2013wja}.

The fermionic Higgs-portal dark matter model could be the SM
with effective interaction~\cite{Kim:2006af}, or renormalizable theory~\cite{Kim:2008pp}.
If we consider another DM component to be fermionic field besides the scalar one, then the number of scalars $N$ is even larger than the one without fermionic component. That is another reason why we choose the vector rather than the fermion
as another DM component. To take vector field as DM candidate, one may use the Higgs-portal vector DM model
directly~\cite{Djouadi:2011aa} or to generalize it to has extra $U(1)$ gauge symmetry thus
to make the model renormalizable~\cite{Baek:2012se,Choi:2013eua}. For the $S$ and $V$ to be stable and may be DM
candidate, we impose $Z_{2}$ symmetry on $S$ and $V$. Therefore the Lagrangian for the vector and scalar DM interacting with SM through the Higgs-portal is
\begin{equation}
  \mathcal{L_{SV}}=\mathcal{L_{S}}+\mathcal{L}_{V}\;,
 \end{equation}
with
 \begin{eqnarray}
 \mathcal{L}_{S}&=&-\frac{m^2_{S}}{2}S^{2}-\frac{1}{4}\lambda_{s}S^{4}
 -\frac{1}{4}\lambda_{hSS}H^{\dag}HS^{2}\;,\nonumber\\
 \mathcal{L}_V&=&\frac{1}{2}m^{2}_{V}V_{\mu}V^{\mu}+\frac{1}{4}\lambda_{V}(V_{\mu}V^{\mu})^{2}\nonumber\\
 &+& \frac{1}{4}\lambda_{hVV}H^{\dag}HV_{\mu}V^{\mu}\;,
 \end{eqnarray}
where the scalar and vector DM mass are given by
\begin{eqnarray}
 &&M^2_{S}=m^2_{s}+\frac{1}{4}\lambda_{hSS}v^2,\nonumber\\
 &&M^{2}_{V}=m^{2}_{V}+\frac{1}{4}\lambda_{hVV}v^{2}\;.
 \end{eqnarray}

\begin{figure}[!htb]
  \includegraphics[width=3in]{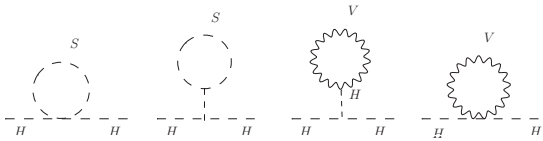}\\
  \caption{Scalar and vector singlet field contributions to %
   the Higgs self-energy.}\label{fig:SVFT}
\end{figure}

In this model, the $VC_{SM}$  becomes
 \begin{eqnarray}
 VC_{SV}&=&6m^{2}_{h}-24m^{2}_{t}+12m^{2}_{W}
 +6m^{2}_{Z}+\frac{\lambda_{hSS}v^{2}}{4}\nonumber\\
 &+& 4\lambda_{hVV}v^{2}\;.
 \label{eq:vssm}
 \end{eqnarray}
 From the Eq.(\ref{eq:vssm}), we found that $VC_{SV}=0$ is easier to satisfy than that of $VC_{SM}=0$, which could also be seen in Fig.~\ref{fig:SVFT}.

The significant difference from the previous paper
~\cite{Grzadkowski:2009mj,Chakraborty:2012rb} is that we
take into account the tadpole contributions as could be seen in Fig.~\ref{fig:SVFT}.

In order not to reintroduce the hierarchy problem, the mass of the $S$ to cure the vacuum
instability should not be far away from the weak scale~\cite{Chun:2013soa}, in the case of
$M_S>M_V$ and that 1 TeV$<M_{S}<10$ TeV, the fine-tuning of $M_{S}$
also need to be considered. It could be used to fix $\lambda_{S}$ in terms of Higgs-portal
couplings $\lambda_{hSS}, \lambda_{hVV}$, both of which will be constrained
by the observed DM relic density and direct detection experiments.

\section{Relic density constraints}

\begin{figure}[!htbp]
\begin{center}
\includegraphics[width=3.0in]{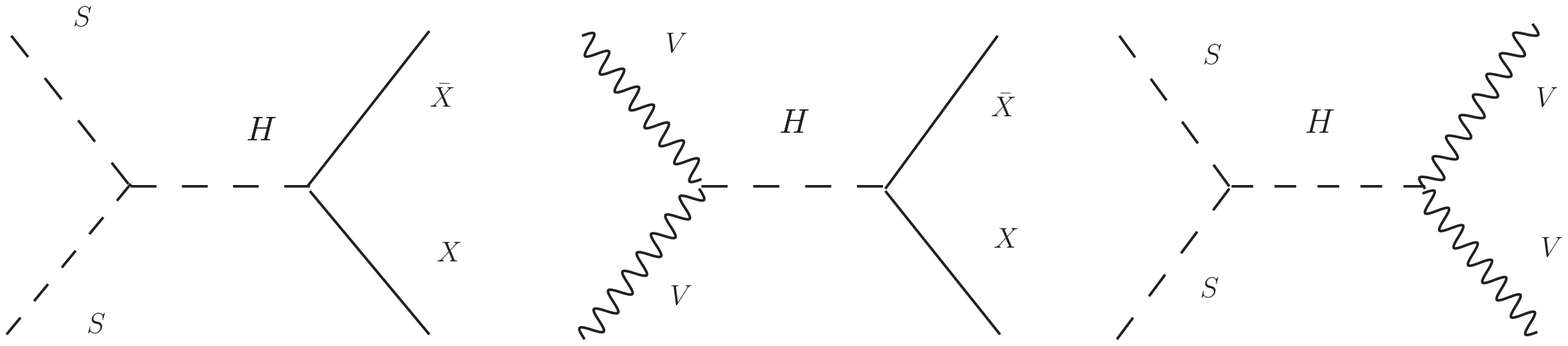}
\end{center}
\caption{Feynman diagrams for dark matter annihilation,
the first two diagrams for the channels of two $S,V$ annihilate into a pair of SM particles $SS\rightarrow X\bar{X}, VV\rightarrow X\bar{X}$, and the last one for the annihilation channels $SS\rightarrow VV, VV\rightarrow SS$.}\label{fig:ddsm}
\end{figure}

In our two component DM model, the DM sector contains a real gauge singlet scalar field, $S$ and a gauge singlet vector field $V$. Both of them will contribute to the total DM relic density in the universe. The leading annihilation channel is that the DM pairs annihilate into SM particles through Higgs exchange, $SS(VV)\rightarrow h^{*}\rightarrow X \bar{X}$, where $X$ ($\bar{X}$) stands for SM particles (anti-particles) such as leptons and quarks, gauge bosons and Higgs boson. The corresponding Feynman diagrams are shown in the first two diagrams in Fig.~\ref{fig:ddsm}. Besides both are annihilating into SM particles, the two DM candidate, $S$ and $V$ can also annihilate into each other which are shown in the last Feynman diagrams in Fig.~\ref{fig:ddsm} (here the initial and final state of the annihilate processes depend on the mass threshold between $M_{S}$ and $M_{V}$).

Following Ref.~\cite{Belanger:2011ww}, the coupled Boltzman equations which govern the evolution of the DM components number density $n_{i,j}$ ($i,j=S,V$) can be written as
\begin{eqnarray}\label{eq:copn}
 \frac{d n_{i}}{d t}+3n_{i} H&=&-\langle\sigma v_{rel}\rangle_{ii\rightarrow X\bar{X}}
 \left(n^{2}_{i}-(n^{eq}_{i})^2\right) \nonumber\\
 &-& \langle\sigma v_{rel}\rangle_{ii\rightarrow jj}
 \left(n^{2}_{i}-\frac{(n^{eq}_{i})^{2}}{(n^{eq}_{j})^2}
 n^{2}_{j}\right) \;,\nonumber\\
 \frac{d n_{j}}{dt}+3n_{j} H&=&-\langle\sigma v_{rel}\rangle_{jj\rightarrow X\bar{X}}
 \left(n^{2}_{j}-(n^{eq}_{j})^2\right) \nonumber\\
 &+& \langle\sigma v_{rel}\rangle_{ii\rightarrow jj}\left(n^{2}_{i}-
 \frac{(n^{eq}_{i})^{2}}{(n^{eq}_{j})^2}n^{2}_{j}\right) \;,
 \end{eqnarray}
where $n^{eq}_{i,j}$ is the equilibrium number density of component $i$, $H$ is the Hubble parameter. It is convenient to introduce two dimensionless variables $Y_{i,j}=\frac{n_{i,j}}{s}$ and $x_{i,j}=\frac{m_{i,j}}{T}$~\cite{Gondolo:1990dk}, where $s$ and $T$ are the entropy density and temperature of the universe. With the new variables, the Boltzman equations can be recast as
\begin{eqnarray}\label{eq:boltzSV}
  \frac{d Y_{S}}{d x_{S}}&=&-
  \frac{1.32g^{1/2}_{\star}M_{S}M_{p}}{x^{2}_{S}}\bigg(\langle\sigma v_{rel}\rangle_{SS\rightarrow X\bar{X}}\big(Y^{2}_{S}-
  (Y^{eq}_{S})^{2}\big)\nonumber\\
  &+&
  \langle\sigma v_{rel}\rangle_{SS\rightarrow VV}
  \bigg(Y^{2}_{S}-\frac{(Y^{eq}_{S})^{2}}{(Y^{eq}_V)^{2}}
  Y^{2}_{V}\bigg)\bigg)\;,\nonumber\\
  \frac{d Y_V}{d x_V}&=&-
  \frac{1.32g^{1/2}_{\star}M_VM_{p}}{x^{2}_V}\bigg(\langle\sigma v_{rel}\rangle_{VV\rightarrow X\bar{X}}\big(Y^{2}_V-
  (Y^{eq}_V)^{2}\big)\nonumber\\
  &-&\langle\sigma v_{rel}\rangle_{SS\rightarrow VV}
  \bigg(Y^{2}_{S}-\frac{(Y^{eq}_{S})^{2}}{(Y^{eq}_V)^{2}}
  Y^{2}_V\bigg)\bigg)\;,
\end{eqnarray}
for $M_{S}>M_{V}$. Similarly, for $M_{V}>M_{S}$, one has
\begin{eqnarray}\label{eq:boltzVS}
  \frac{d Y_V}{d x_V}&=&-
  \frac{1.32g^{1/2}_{\star}M_V M_{p}}{x^{2}_V}\bigg(\langle\sigma v_{rel}\rangle_{VV\rightarrow X\bar{X}}\big(Y^{2}_V-
  (Y^{eq}_V)^{2}\big)\nonumber\\
  &+& \langle\sigma v_{rel}\rangle_{VV\rightarrow SS}
  \bigg(Y^{2}_{V}-\frac{(Y^{eq}_{V})^{2}}{(Y^{eq}_S)^{2}}Y^{2}_S
  \bigg)\bigg)\;,\nonumber\\
  \frac{d Y_{S}}{d x_{S}}&=&-
  \frac{1.32g^{1/2}_{\star}M_{S}M_{p}}{x^{2}_{S}}\bigg(\langle\sigma v_{rel}\rangle_{SS\rightarrow X\bar{X}}\big(Y^{2}_{S}-
  (Y^{eq}_{S})^{2}\big)\nonumber\\
  &-& \langle\sigma v_{rel}\rangle_{VV\rightarrow SS}
  \big(Y^{2}_{V}-\frac{(Y^{eq}_{V})^{2}}{(Y^{eq}_S)^{2}}
  Y^{2}_S\bigg)\bigg)\;.
\end{eqnarray}
Here $M_{p}=2.44\times10^{18}$ GeV is the reduced Planck mass, and $g_{\star}$ is the degrees of freedom parameter. Solving the coupled Eqs.~(\ref{eq:boltzSV},\ref{eq:boltzVS}), we get the values of $Y_{S}$ and $Y_{V}$ at present temperature $T_{0}$.
As we discussed previously, $\langle\sigma v\rangle_{SS (VV)\rightarrow X\bar{X}}$ in Eqs.~(\ref{eq:boltzSV},\ref{eq:boltzVS}) represents the total annihilation cross sections of DM particles annihilating into SM particles through Higgs boson exchange, i.e., $SS(VV)\rightarrow h^{*}\rightarrow X \bar{X}$. The corresponding expressions of these cross sections are given below
\begin{eqnarray}\label{eq:sigs}
  \langle\sigma v_{rel}\rangle_{SS\rightarrow X \bar{X}}&=&\frac{\lambda_{hSS}^2v^2/2}
  {(4M_{S}^2-m_{h}^2)^2+\Gamma_{h}^2 m_{h}^2}\nonumber\\
  &\times& \frac{\sum_{i}\Gamma{(\tilde{h}
  \rightarrow{X_{i}})}}{2M_{S}}\;,\nonumber\\
 \langle\sigma v_{rel}\rangle_{VV\rightarrow X \bar{X}}&=&\frac{2\lambda_{hVV}^2v^2}
 {(4M_{V}^2-m_{h}^2)^2+\Gamma_{h}^2 m_{h}^2}\nonumber\\
 &\times& \frac{\sum_{i}\Gamma{(\tilde{h}
  \rightarrow{X_{i}})}}{2M_{V}}\;.
\end{eqnarray}
In the above equations, $v_{rel}=2|\mathbf{p}_{S,V}^{cm}|/M_{S,V}$ is the relative
velocity of the two DM particles in center-of-mass (c.m.s) frame, $\tilde{h}$ is
the virtaul Higgs boson with an invariant mass $\sqrt{s}=2M_i$,
$\sum_{i}\Gamma(\tilde{h}\rightarrow{X_{i}})$ represents its total decay width, where the sum runs over all possible decay mode for $\tilde{h}$ into SM particles except the Higgs boson, $\Gamma_{h}$ the total decay width corresponding to the Higgs boson mass $m_{h}=125 \GeV$. In this work, we calculate $\sum_{i}\Gamma(\tilde{h}\rightarrow{X_{i}})$ and $\Gamma_{h}$ with program {\tt hdecay}~\cite{Djouadi:1997yw}. Furthermore, when DM mass $M_{S,V}$ is smaller than Higgs boson mass $m_{h}$, the contributions from
 decay channels
\begin{eqnarray}
 \Gamma(h\rightarrow V V)&=&\frac{\lambda^{2}_{hVV}v^{2}m^{3}_{h}
 \sqrt{1-4M^{2}_{V}/m^{2}_{h}}}{128\pi M^{4}_{V}}\nonumber\\
 &\times&\bigg(1-
 4\frac{M^{2}_{V}}{m^{2}_{h}}+12\frac{M^{4}_{V}}{m^{4}_{h}}
 \bigg)\;,\nonumber\\
 \Gamma(h\rightarrow S S)&=&\frac{\lambda^{2}_{hSS}v^{2}}
 {128\pi m_{H}}\left(1-\frac{4M^{2}_{S}}{m^{2}_{h}}\right)\;,
\end{eqnarray}
should be added, while for $M_{S,V}>m_{h}$, annihilation channels
\begin{eqnarray}\label{eq:Vdecay}
 \langle\sigma v_{rel}\rangle_{SS\rightarrow hh}&=&\frac{\lambda^{2}_{hss}}{128\pi M^{2}_{S}}
 \sqrt{1-\frac{m^{2}_{h}}{M^{2}_{S}}},\nonumber\\
 \langle\sigma v_{rel}\rangle_{VV\rightarrow hh}
 &=&\frac{\lambda^{2}_{hVV}}{32\pi M^{2}_{V}}\sqrt{1-m^{2}_{h}/M^{2}_{V}},
\end{eqnarray}
 also contribute to the total annihilation cross sections. Finally, since there exist interactions between two DM components $S$ and $V$, which affect the evolution of DM abundances, the cross sections $\langle\sigma v_{rel}\rangle_{SS\rightarrow VV}$ and $\langle\sigma v_{rel}\rangle_{VV\rightarrow SS}$ need to be taken into account in the coupled Boltzman equations. The corresponding expressions are given by
\begin{eqnarray}\label{eq:coa}
 \langle\sigma v_{rel}\rangle_{SS\rightarrow VV}
 &=&\frac{\lambda^{2}_{hSS}v^2}{32}
 \frac{1}{(4M^{2}_{S}-m^{2}_{h})^{2}+
 \Gamma^{2}_{h}m^{2}_{h}}\nonumber\\
 &\times& \frac{\Gamma(\tilde{h}
 \rightarrow V V)}{2M_{S}}\;,
 \end{eqnarray}
for mass mass threshold $M_{S}>M_{V}$, and
\begin{eqnarray}
 \langle\sigma v_{rel}\rangle_{VV\rightarrow SS}&=&\frac{2\lambda_{hVV}^2v^2}
 {(4M_{V}^2-m_{h}^2)^2+\Gamma_{h}^2 m_{h}^2}\nonumber\\
 &\times&\frac{\Gamma{(\tilde{h}
 \rightarrow{SS})}}{2M_{V}}\;,
\end{eqnarray}
for $M_{V}>M_{S}$.

\begin{figure}[!htbp]
\begin{center}
\includegraphics[width=0.49\linewidth]{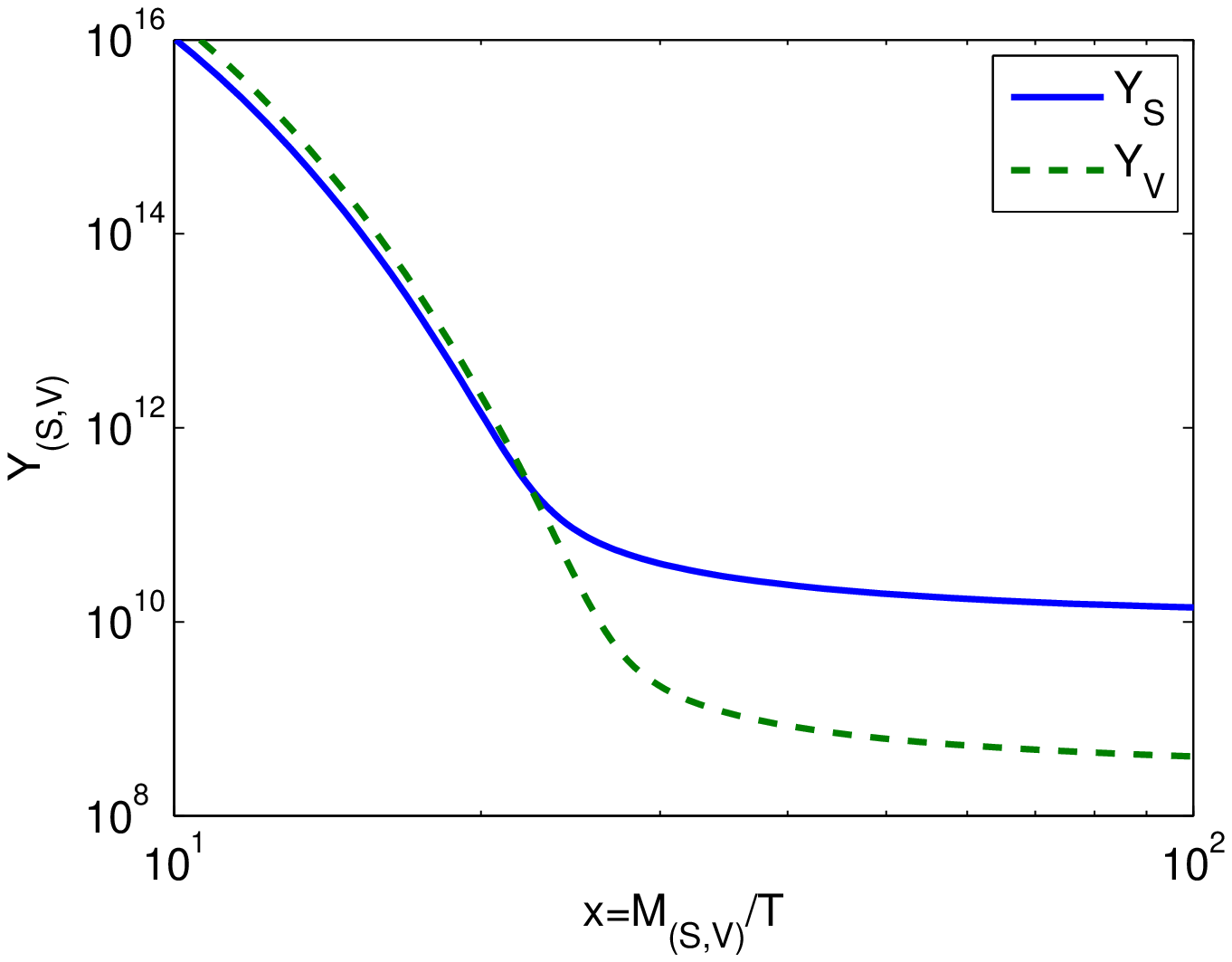}
\includegraphics[width=0.49\linewidth]{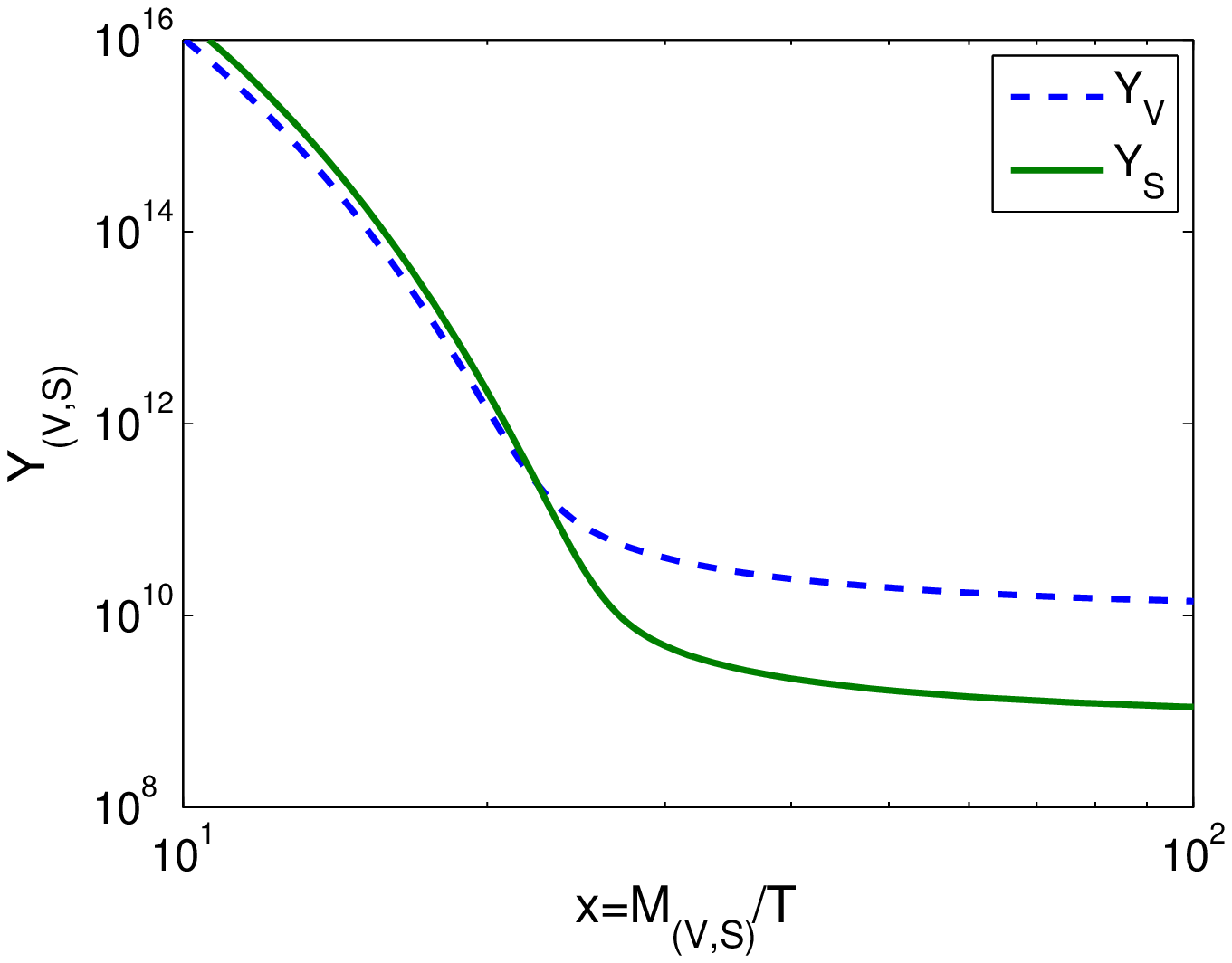}
\includegraphics[width=0.49\linewidth]{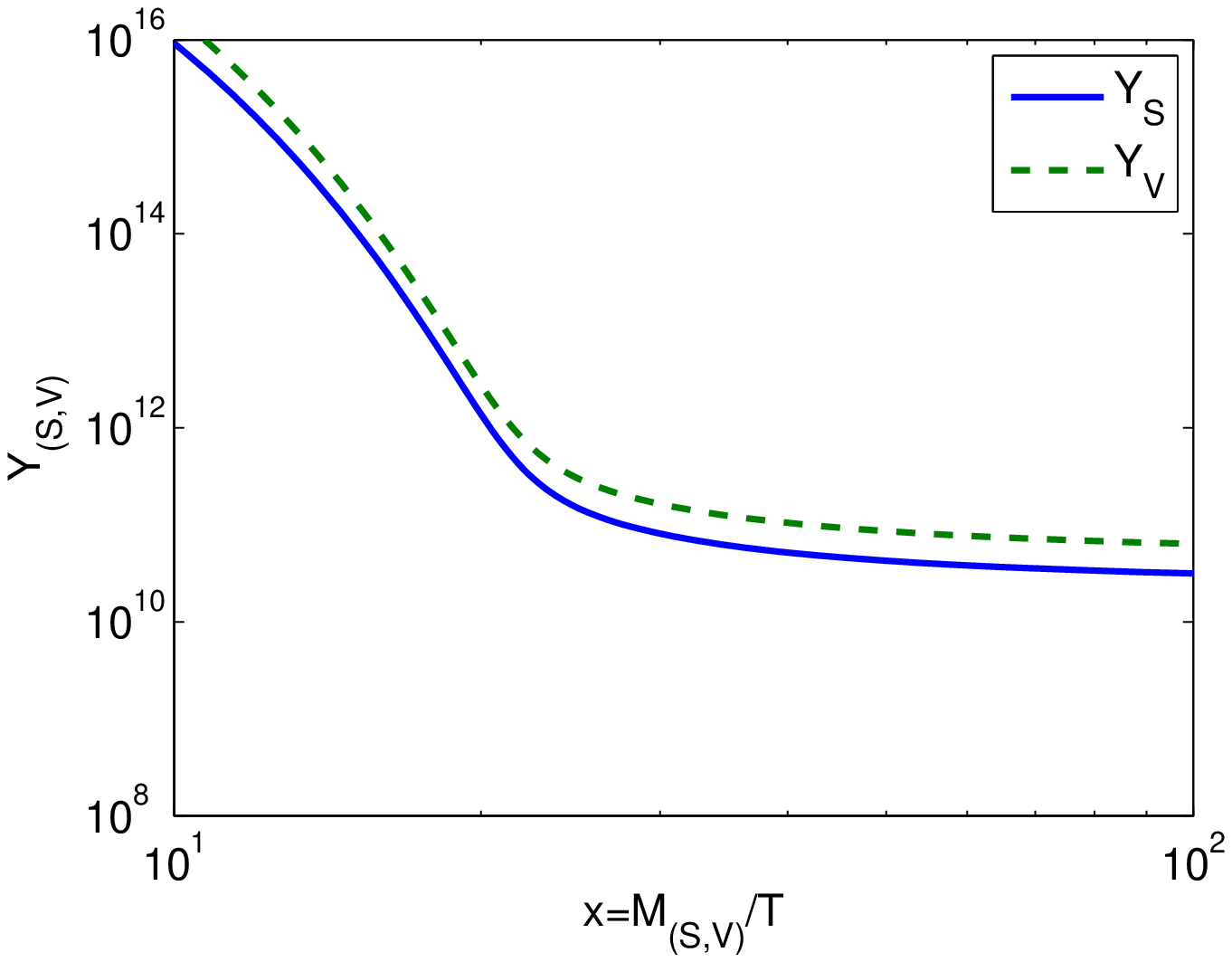}
\end{center}
\caption{The evolution of the abundances of $S$ (solid line) and
$V$ (dashed line) as a function of $x_{S,V} \equiv
M_{S,V}/T$. DM mass are fixed as $M_{S}=160$ GeV, $M_{V}=130$ GeV (top-left), $M_{S}=130$ GeV, $M_{V}=160$ GeV (top-right) and  $M_{S}=160$ GeV, $M_{V}=16$ GeV (bottom), respectively. \label{fig:abundances}}
\end{figure}

\begin{figure}[!htbp]
\begin{center}
\includegraphics[width=0.49\linewidth]{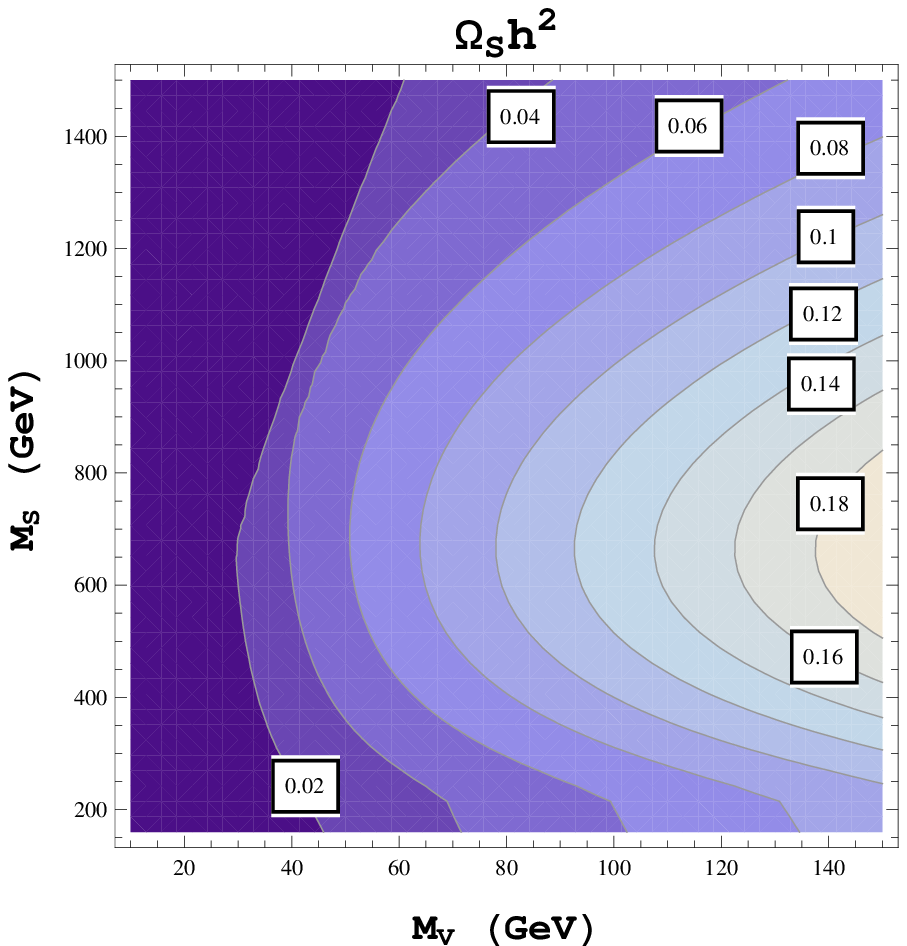}
\includegraphics[width=0.49\linewidth]{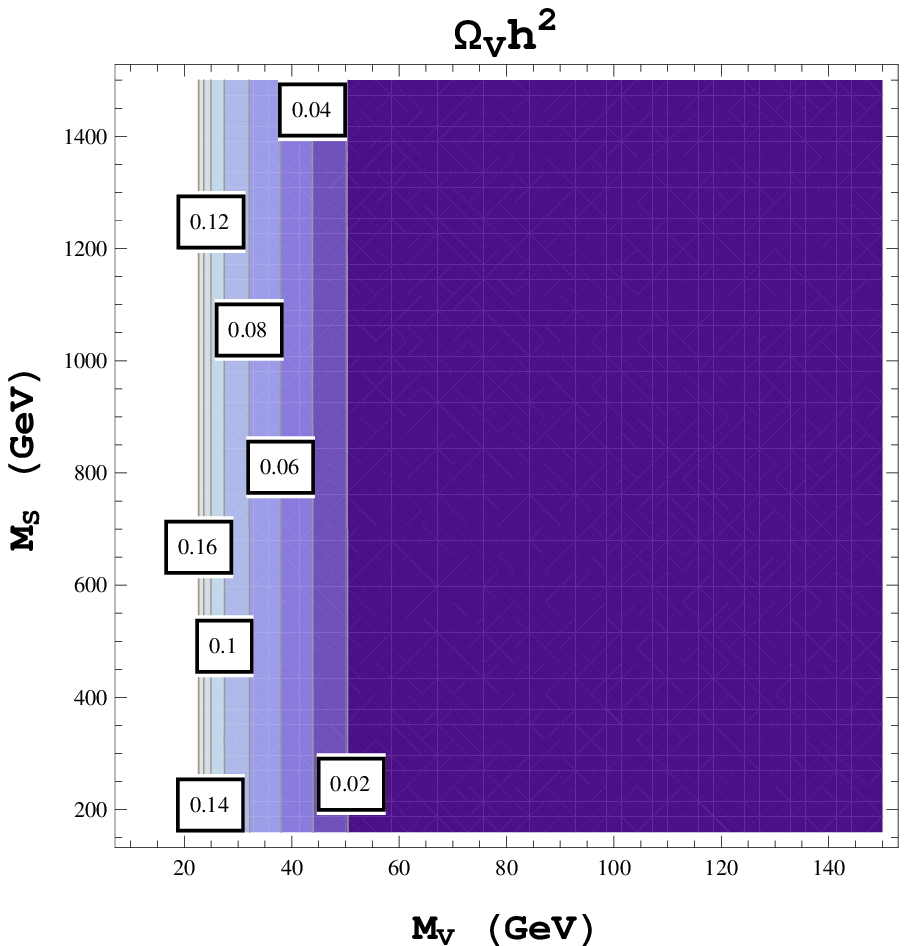}
\includegraphics[width=0.49\linewidth]{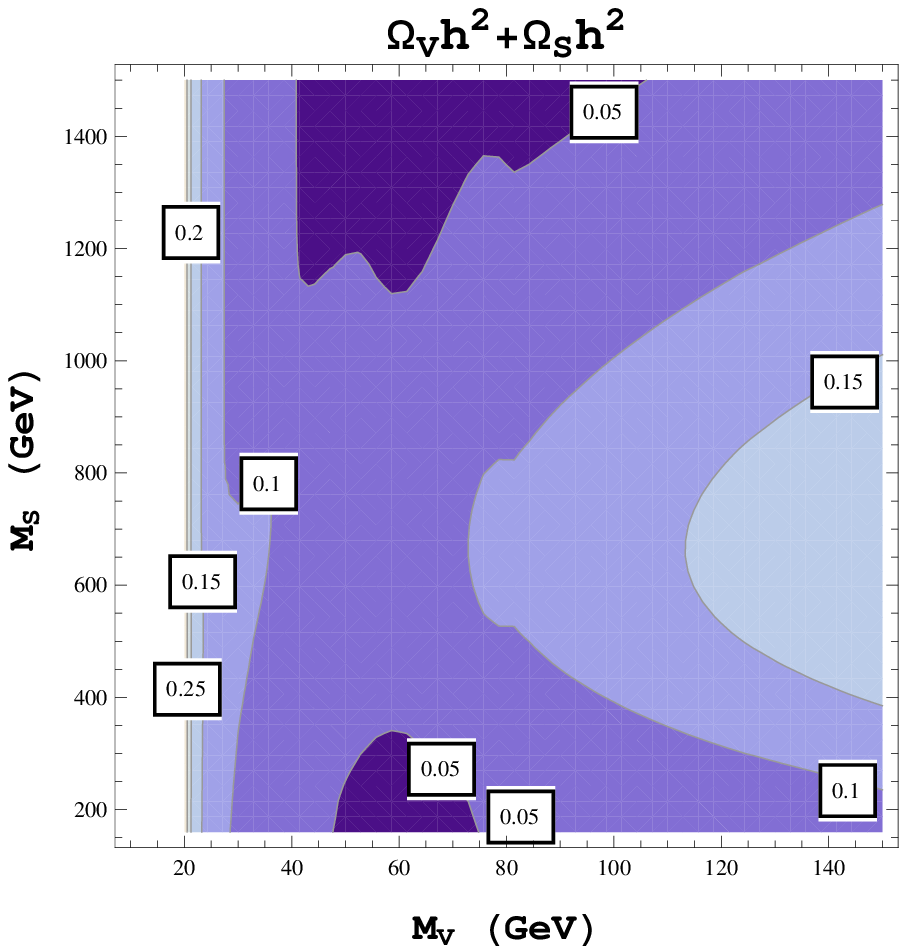}
\end{center}
\caption{Contour plots of relic density in the $M_{S}-M_{V}$ plane for the case $M_{S}>M_{V}$. Where the top-left (top-right) panel corresponds to scalar (vector) component, respectively.
The total relic density $\Omega_{S}h^{2}+\Omega_{V}h^{2}$ is shown in the bottom panel. Here we fix the coupling parameters with $\lambda_{hSS}=0.1$ and $\lambda_{hVV}=0.3$. \label{fig:relicsv1}}
\end{figure}

\begin{figure}[!htbp]
\begin{center}
\includegraphics[width=0.49\linewidth]{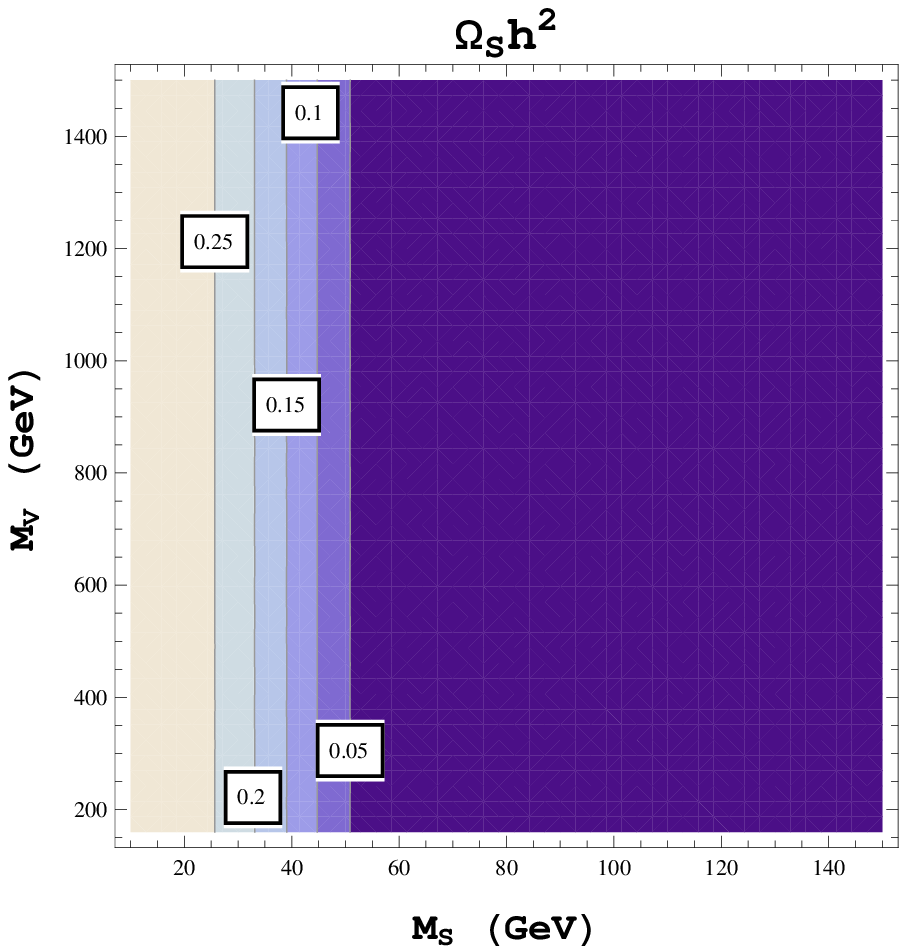}
\includegraphics[width=0.49\linewidth]{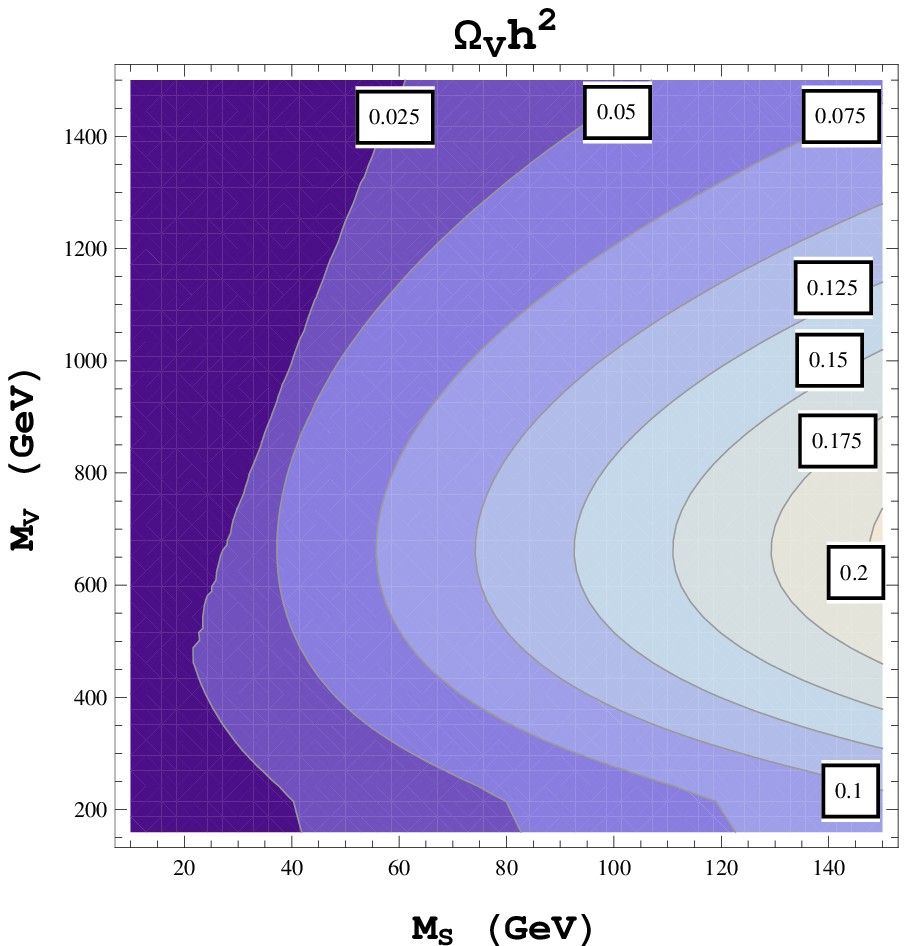}
\includegraphics[width=0.49\linewidth]{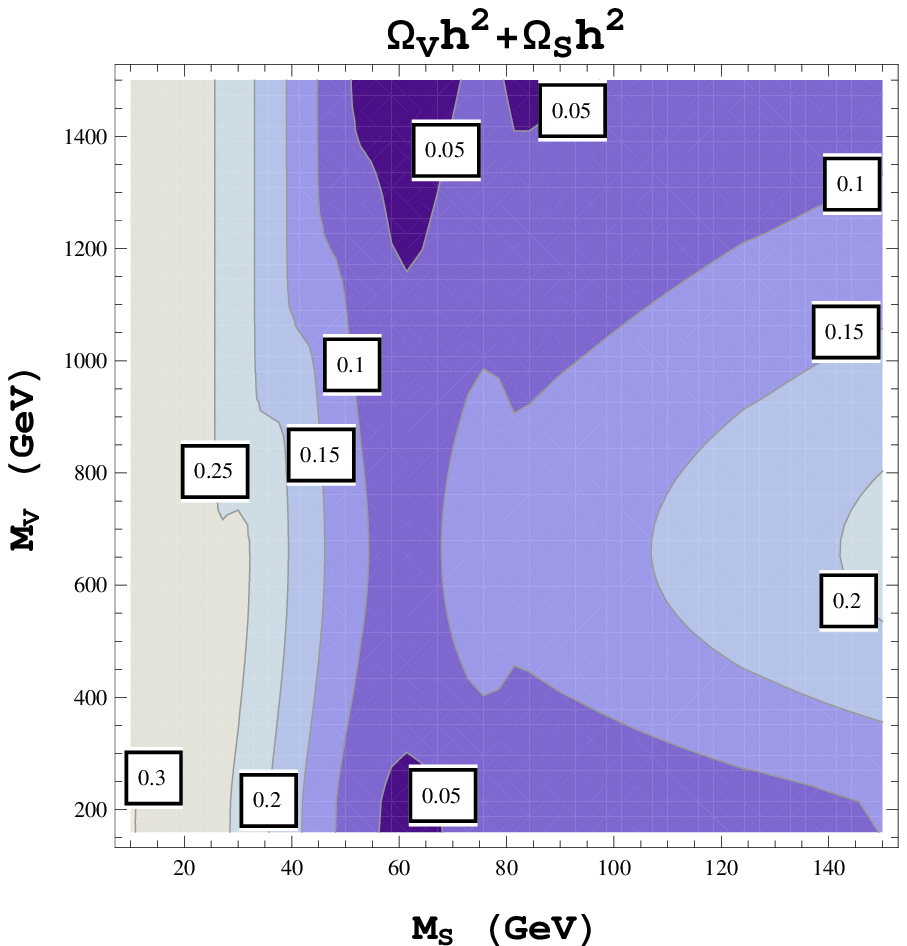}
\end{center}
\caption{Similar to Fig.~\ref{fig:relicsv1}, but for the case $M_{S}<M_{V}$. Coupling parameters are fixed as $\lambda_{hSS}=0.35$ and $\lambda_{hVV}=0.05$. \label{fig:relicsv2}}
\end{figure}

\begin{figure}[!htbp]
\begin{center}
\includegraphics[width=0.49\linewidth]{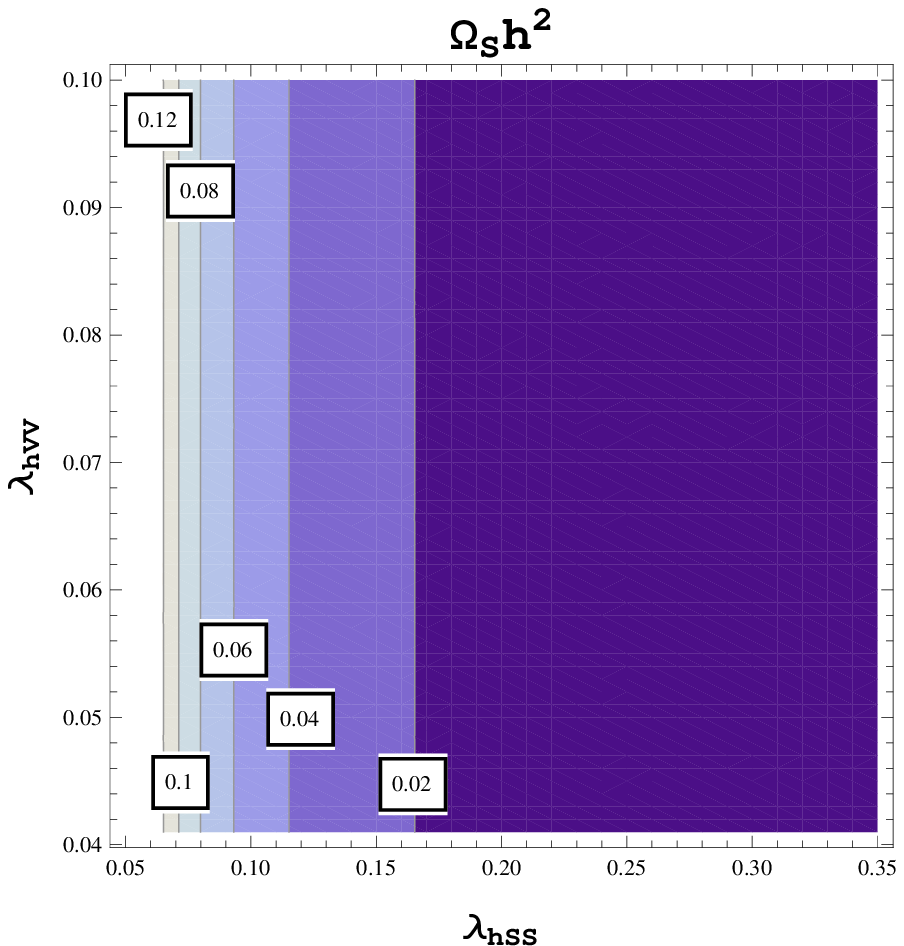}
\includegraphics[width=0.49\linewidth]{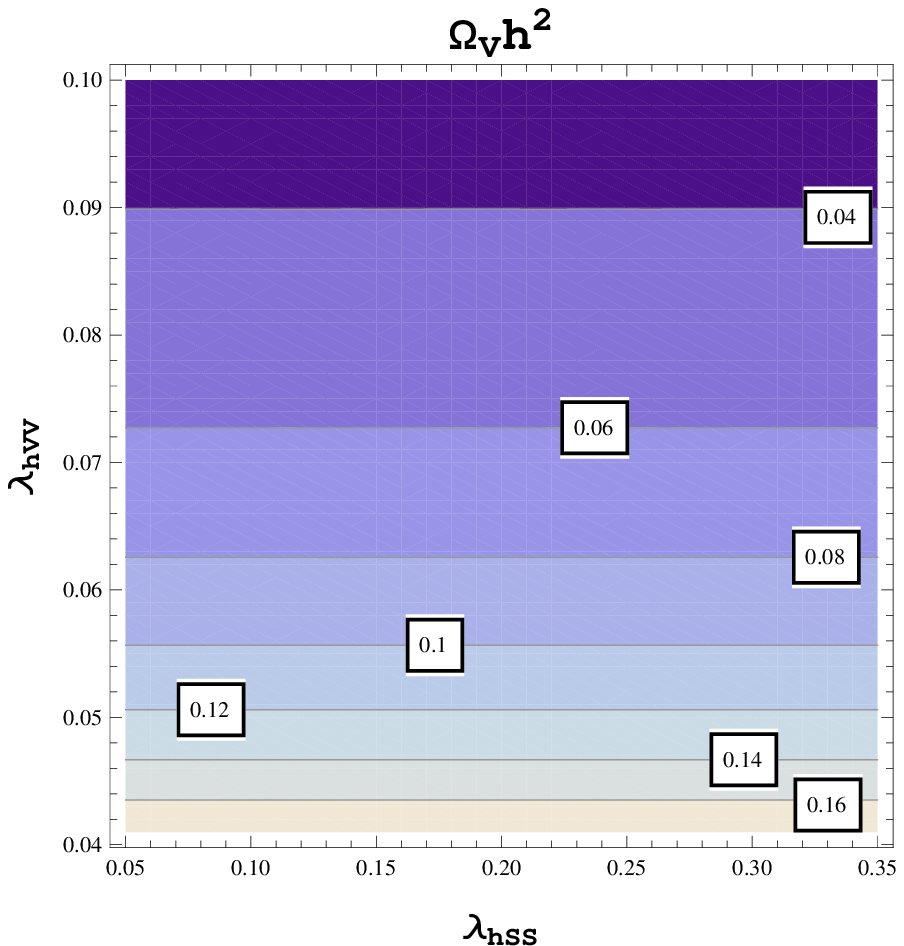}
\includegraphics[width=0.49\linewidth]{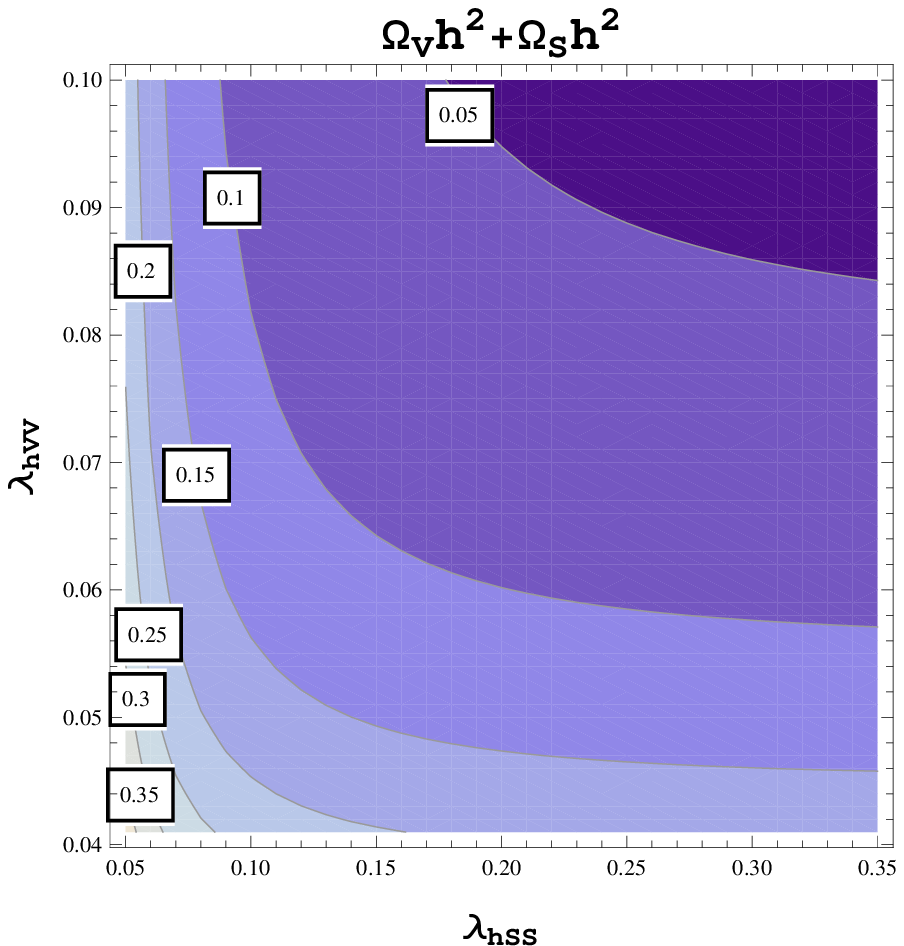}
\end{center}
\caption{Contour plots for the relic density of the DM particles
$S$ (top left), $V$ (top right) and total relic density (bottom)in the $\lambda_{hSS}-\lambda_{hVV}$ plane with the fixed mass parameters $M_V$=500 GeV and $M_S$=100 GeV. \label{fig:relicsv3}}
\end{figure}

With all the cross sections given above, we solve the coupled Eqs. (\ref{eq:boltzSV},\ref{eq:boltzVS}) numerically
to obtain the abundance of $S$ and $V$. The relic density for each component can be calculated through
$\Omega_{S,V}h^{2}=2.755\times10^{8}\frac{M_{S,V}}{GeV}Y_{S,V}(T_{0})$ ~\cite{Edsjo:1997bg,Biswas:2013nn}, and the total relic density of DM is the sum of two components, $\Omega h^{2}=\Omega_{S}h^{2}+\Omega_{V}h^{2}$. There are four free parameters in our model, including two DM mass parameters $M_{S},M_{V}$ as well as two coupling parameters $\lambda_{hSS}$ and $\lambda_{hVV}$. To illustrate how the interplay between the two DM components affects the evolution
of DM abundances, we examine the evolution of DM abundances $Y_{S}$ and $Y_{V}$ in two limits:
\begin{enumerate}
   \item {Interactions between $S$ and $V$ are weaker than with SM particles ( $\langle\sigma v_{rel}\rangle_{SS,VV\rightarrow VV,SS}\\\ll\langle\sigma v_{rel}\rangle_{SS,VV\rightarrow X \bar{X}}$ ).
   In this limit, we take parameters as $M_{S}=160$ (130) GeV, $M_{V}=130$ (160) GeV, $\lambda_{hSS}=0.1$ (0.35), and
   $\lambda_{hVV}=0.3$ (0.05) for $M_{S}>M_{V}$ ($M_{S}<M_{V}$).}
   \item {Interactions between $S$ and $V$ are stronger than with SM particles ( $\langle\sigma v_{rel}\rangle_{SS,VV\rightarrow VV,SS}\\\gg\langle\sigma v_{rel}\rangle_{SS,VV\rightarrow X \bar{X}}$ ). In this limit, the parameters are taken to be as $M_{S}=160$ GeV, $M_{V}=16$ GeV, $\lambda_{hSS}=0.1$, and $\lambda_{hVV}=0.3$.}
  \end{enumerate}
The DM abundances $Y_{S,V}$ as a function of $x_{S,V}$ for both two limits are shown in Fig.~\ref{fig:abundances}. Here, one can see that the annihilation between two DM components imprints on the evolution of abundances. In the first limit, DM abundances evolve almost independently, and the final DM abundances are mainly determined by their interaction with SM particles. While in the second limit, since the heavier DM component annihilate into lighter one, the abundance of lighter DM component increases in comparison with the previous case. Thus the final total relic density mainly contributed from lighter DM component.
%
%
%
%
In our model, both annihilation between $S,V$ pairs and $S,V$ annihilate into SM particles proceed via $s$-channel exchange of Higgs boson.
The dominant annihilation channel are determined by the DM mass $M_{S}, M_{V}$ and coupling parameters $\lambda_{hSS}, \lambda_{hVV}$. The $s-$channel exchange can be
strongly enhanced by a resonance effect. We therefore expect the
relic density to drop rapidly when $M_{S}, M_{V} \approx m_h/2$.
 In order to investigate the
dependence of the DM relic density on each
parameter, we scan the parameter space according to the following three groups:
 \begin{enumerate}
   \item {Scan in the $M_{S} - M_{V}$ plane for the case $M_{S}>M_{V}$, where we choose  coupling parameters as $\lambda_{hSS}=0.1$, $\lambda_{hVV}=0.3$.}
   \item {Scan in the $M_{S} - M_{V}$ plane for the case $M_{S}<M_{V}$, and we choose  coupling parameters as $\lambda_{hSS}=0.35$, $\lambda_{hVV}=0.05$.}
   \item {Scan in the $\lambda_{hSS} - \lambda_{hVV}$ plane, here we fix DM mass with $M_V$=500 GeV, $M_S$=100 GeV.}
 \end{enumerate}
Contour plots for three groups parameter space discussed above are displayed in Figs.~\ref{fig:relicsv1} - \ref{fig:relicsv3}, respectively. Where the relic density for each component are shown in the top panel, and the total relic density $\Omega_{S}h^2 + \Omega_{V}h^2$ are shown in the bottom panel. Next, we will analysis them in detail one by one.

For the first group parameter space (see Fig.~\ref{fig:relicsv1}), we first note that the enhanced annihilation near the $m_h/2$ resonances explains the decrease in total DM relic density for $M_{V} \approx 60$ GeV. Furthermore, the behavior of DM relic density for each component depends on the mass of vector component $M_{V}$ much heavier than the scalar one. In the large regions of $M_{V}$ ( $M_{V}> 80$ GeV ), the DM relic density is dominated by $S$ while $\Omega_{V}h^2$ dominates for $M_{V}\approx20-30$ GeV. In the intermediate regions of $M_{V}$, both components give a significant contribution to the total DM relic density. This behavior can be explained as follows: the annihilation process $SS\rightarrow VV$ plays a crucial role in this case, the cross section $\langle\sigma v_{rel}\rangle_{SS\rightarrow VV}$ give the significant contribution when the $M_{S}$ and $M_{V}$ are both small, while it is negligible compared with $\langle\sigma v_{rel}\rangle_{SS,VV\rightarrow X \bar{X}}$ when $M_{S}$ and $M_{V}$ are large. Thus, in the large mass region, the DM relic density are mainly determined by
their interaction with SM particles and due to the coupling parameters which we are chosen the final DM abundance is dominated with $S$. However, in the small mass region, the abundance of $V$ get a considerable increase through annihilation process $SS\rightarrow VV$ and dominates over that of $S$.

For the second group, i.e., $M_{V}>M_{S}$ (Fig.~\ref{fig:relicsv2}), the patterns of contour are similar with Fig.~\ref{fig:relicsv1}. While for now, the behavior of $S$ and $V$ are interchanged since the annihilation process becomes $\langle\sigma v_{rel}\rangle_{VV\rightarrow SS}$.

Finally, we examine the dependence of the DM relic density on
these coupling parameters. Here, for simplicity, we choose $M_{V}>M_{S}$ and fix DM mass $M_V$=500 GeV, $M_S$=100 GeV. As was shown by Fig.~\ref{fig:relicsv3}, we found that in this case, the DM relic density for each component almost depend on their own coupling parameters.
This feature is easy to understand, since all annihilation cross sections are proportional to $\lambda_{hSS}$ and $\lambda_{hVV}$.
%
%
%
%
%
%
%
%
\subsection{Direct detection constraints with XENON100}

We have shown the effect of interactions between two DM component on the evolution of DM abundances and on the final DM relic density. In this section, we present the combined constraints based on DM relic density and direct detection experiments. The direct detection of DM measures the event rate and energy
deposit in the scattering of target nuclei by DM particles
in the local galactic halo. In our model, the singlet scalar $S$ and vector $V$ DM interact with the SM particles only through the exchange of the Higgs boson, the DM-nucleon scattering cross section is therefore necessarily spin-independent (SI). For each single component, SI DM-nucleon  cross sections can be easily calculated, which are given by~\cite{Kanemura:2010sh}
\begin{eqnarray}
 \sigma^{S}_{SI}&=&\frac{\lambda^{2}_{hSS}}{16\pi m^{4}_{h}}
 \frac{m^{4}_{N}f^{2}_{N}}{(M_{S}+m_{N})^{2}},\nonumber\\
 \sigma^{V}_{SI}&=&\frac{\lambda^{2}_{hVV}}{16\pi m^{4}_{h}}
 \frac{m^{4}_{N}f^{2}_{N}}{(M_{V}+m_{N})^{2}}.
\end{eqnarray}
Where $m_{N}$ is the nucleon mass and
$f_{N}=\sum f_{L}+3\times\frac{2}{27}f_{H}$ is the effective Higgs-nucleon coupling, which sum the contributions of the light quark ($f_{L}$) and heavy quark ($f_{H}$).
In this work, we use the value in Ref.\cite{Young:2009zb}: $f_{N}=0.326$. Since the current experimental constraints assume that the local DM density is only provided by one DM species. However, in our model the $S,~V$ component all contribute to the local DM density. With the assumption that the contribution of each DM component to the local density is same as their contribution to the relic density, the SI scattering cross section should be rescaled by ${\Omega_{S,V}h^{2}}/{\Omega_{DM}h^{2}}$. Therefore, the corresponding upper limit obeys $\sigma^{S,V}_{SI}\leq ({\Omega_{DM}h^{2}}/{\Omega_{S,V}h^{2}})\sigma^{exp}_{SI}$~\cite{Belanger:2011ww,barger:2008jx}.

\begin{figure}[!htbp]
\begin{center}
\includegraphics[width=0.49\linewidth]{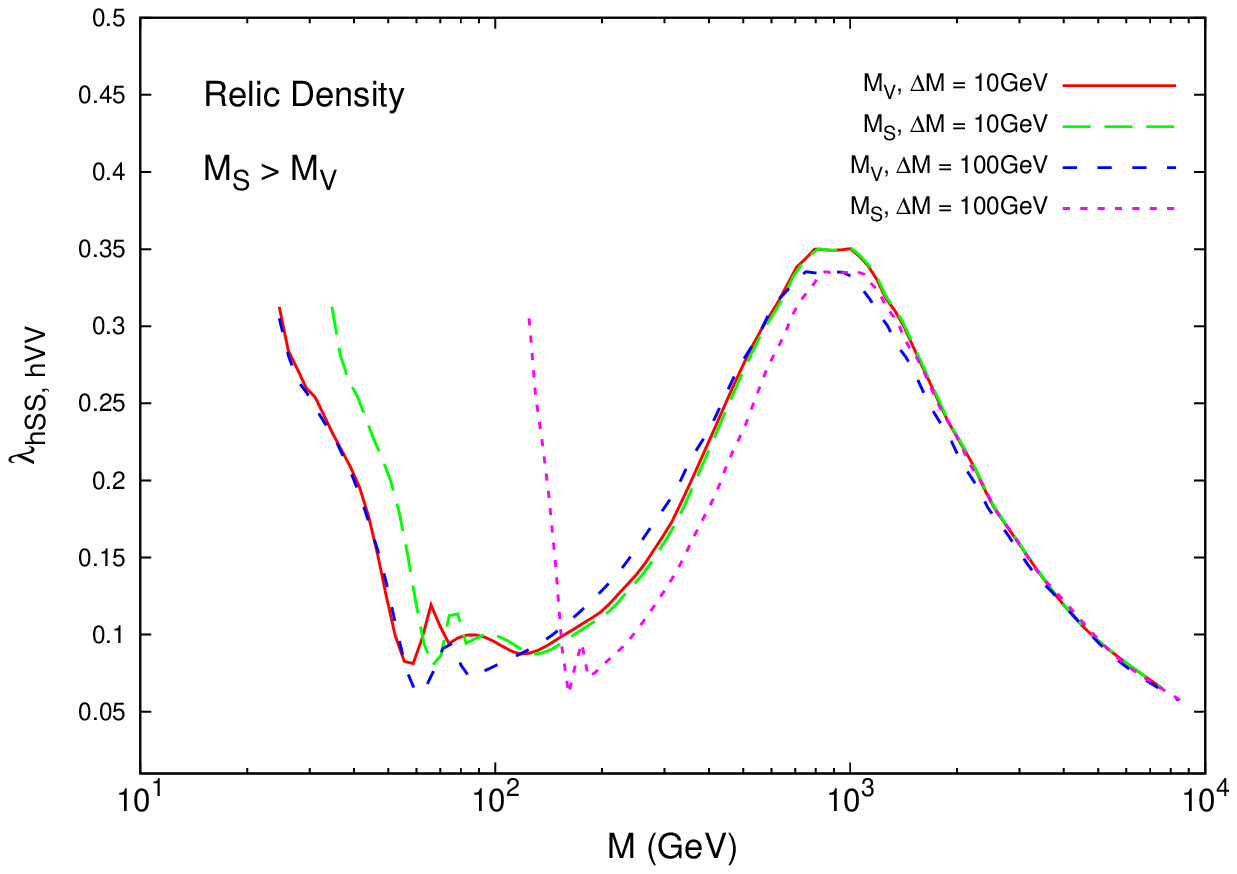}%
\includegraphics[width=0.49\linewidth]{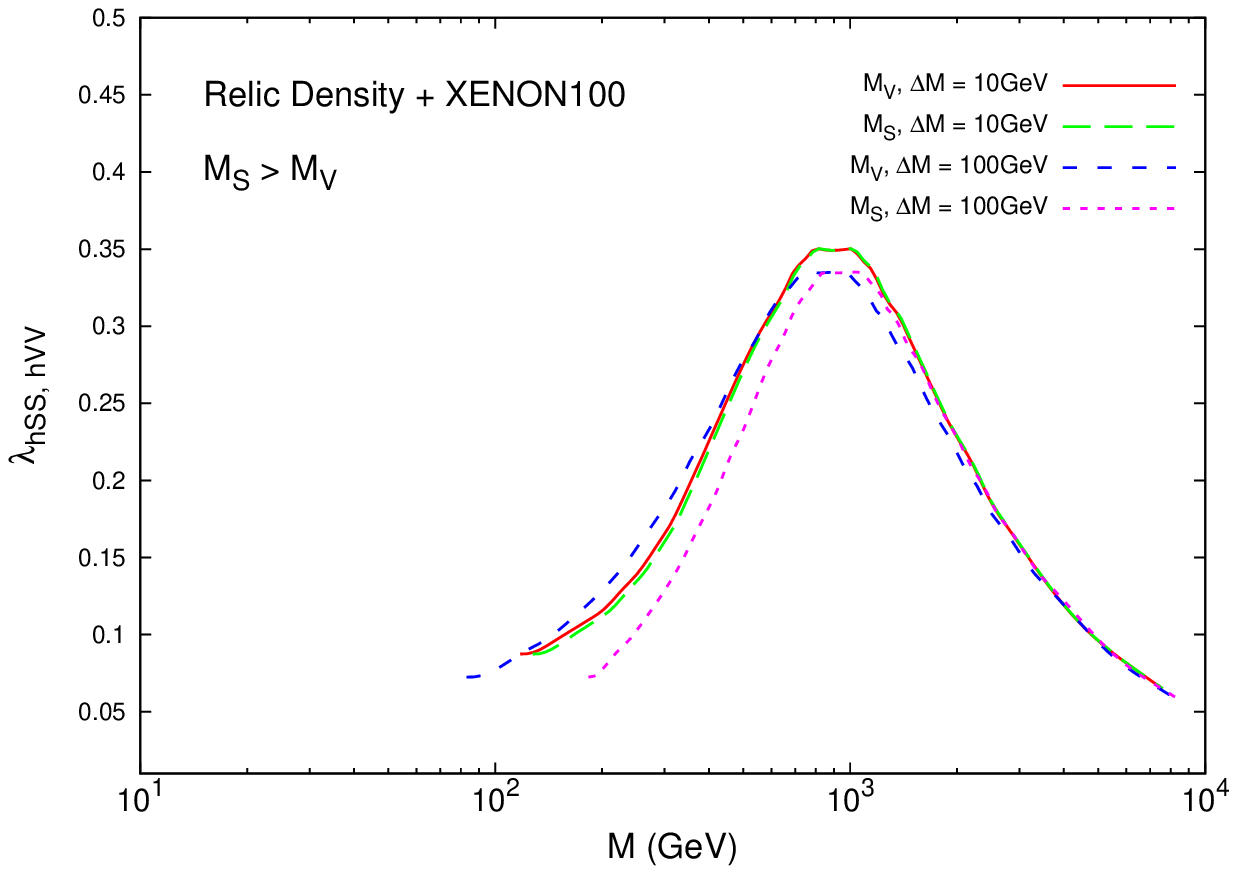}%
\end{center}
\caption{Allowed parameter regions in the $\lambda_{hSS,hVV}-M_{S}$ plane for the case $M_{S}>M_{V}$.
Left panel present the parameter regions allowed by relic density constraint, and right panel
present the parameter regions allowed by both relic density and XENON100
constraints.%
\label{fig:reldlm1}}
\end{figure}

\begin{figure}[!htbp]
\begin{center}
\includegraphics[width=0.49\linewidth]{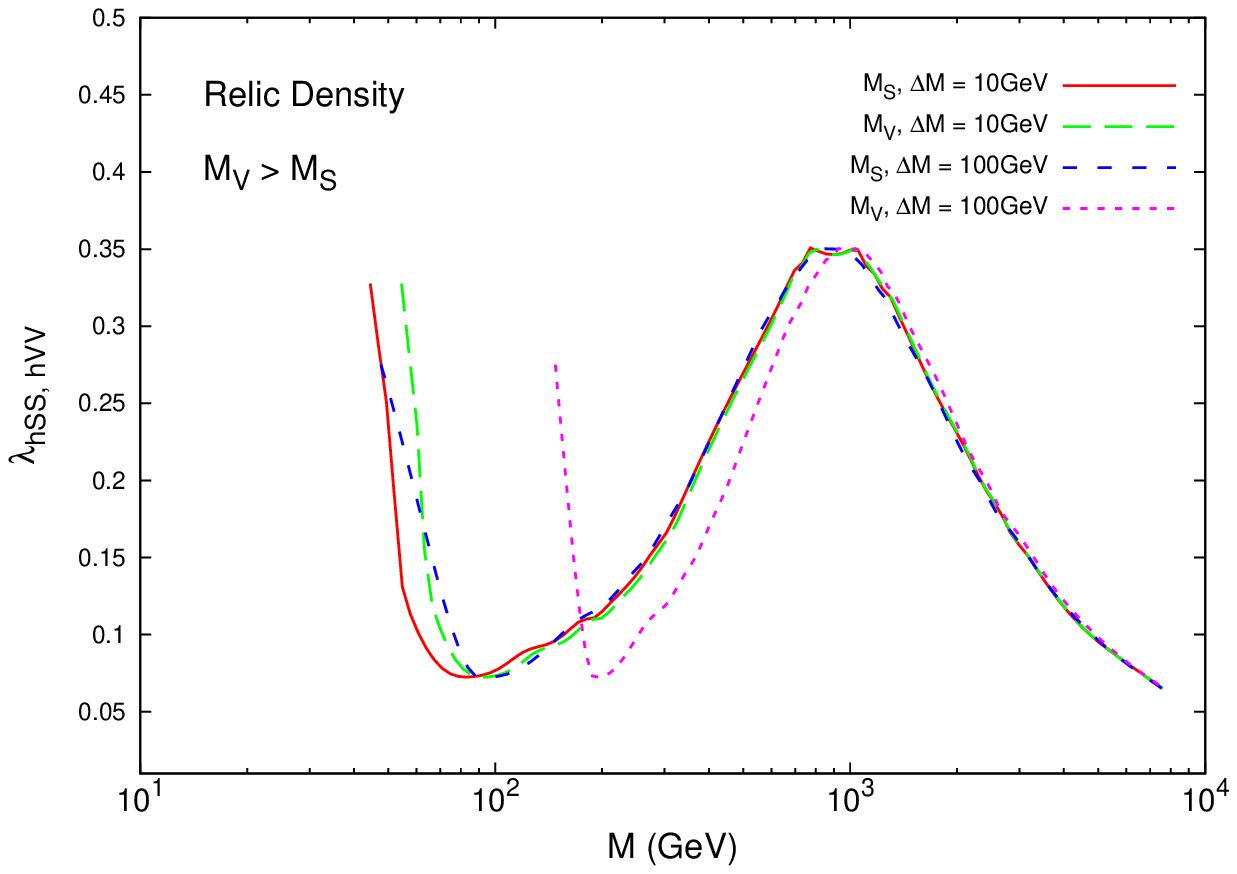}
\includegraphics[width=0.49\linewidth]{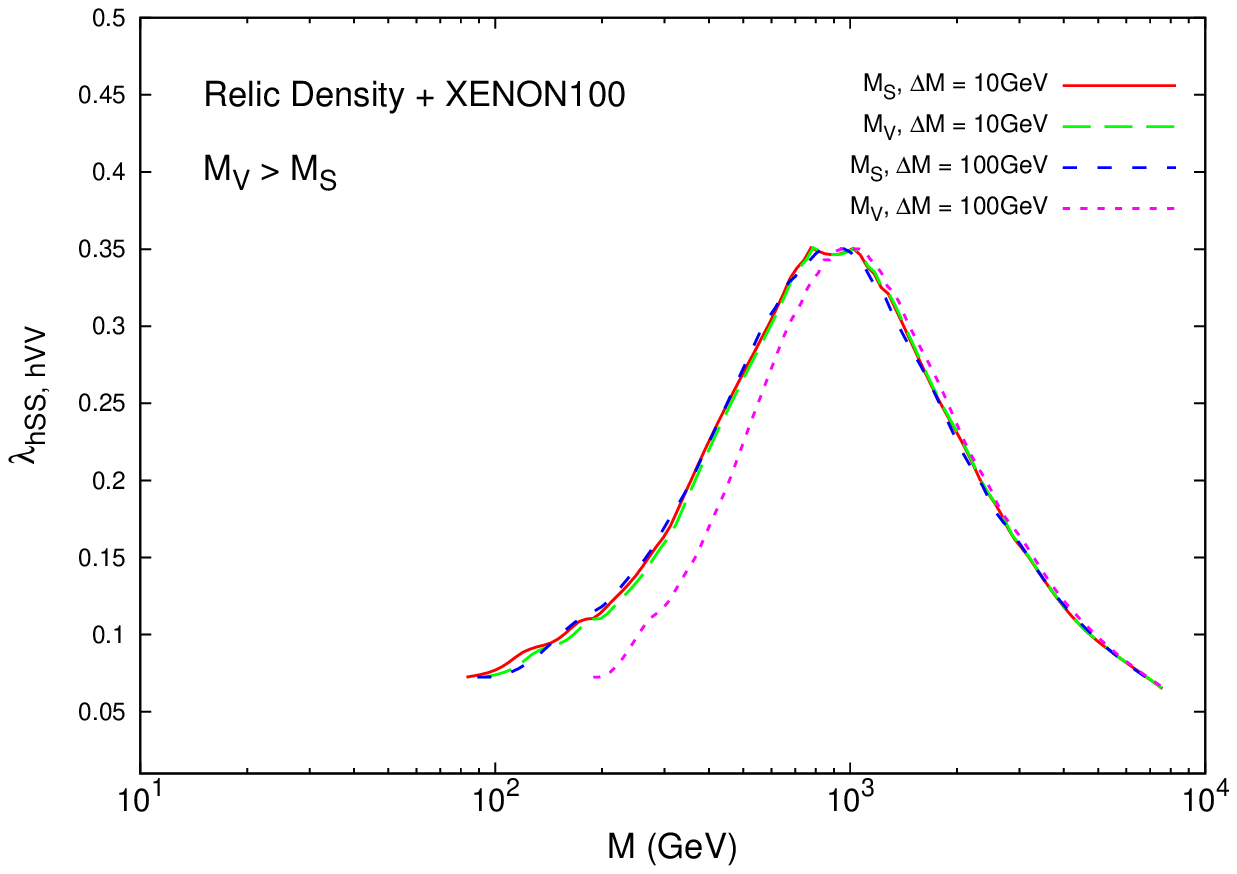}\\
\end{center}
\caption{Similar to Fig.~\ref{fig:reldlm1}, but for the case $M_{S}<M_{V}$ .
\label{fig:reldlm2}}
\end{figure}

In Fig.\ref{fig:reldlm1} and \ref{fig:reldlm2}, we present the parameter space in the
$\lambda_{hSS,hVV}-M_{S,V}$ plane which satisfies the DM relic density and direct detection constraints. For observed relic density, we use the latest Planck+WP+highL+BAO result within $1\sigma$ uncertainty: $0.1187\pm0.0017$\cite{Ade:2013zuv}. For direct detection experiment, we use the recent update result from XENON100 experiment which extends the exclusion region of DM mass up to $10$ TeV \cite{Aprile:2012nq}. To reduce the number of unknown parameters, we simplify the parameter space as follows,\begin{enumerate}
          \item {we set $\lambda_{hSS}=\lambda_{hVV}$.}
          \item {we define the mass difference $\Delta M \equiv M_{S}-M_{V}$ ($M_{V}-M_{S}$) for $M_{S}>M_{V}$ ($M_{V}<M_{S}$) and fix their values with $\Delta M = 10, 100$ GeV, respectively.}
        \end{enumerate}
 In these figures, the left panel present the parameter regions allowed by relic density constraint, and the right panel present the parameter regions allowed by both relic density and XENON100
constraints. As can be seen from the figure, the parameter space are severely constrained by the observed relic density, and the XENON100 limit further exclude the small DM mass region.
It should be noted
there exist tensions among current experimental results. The data from DAMA~\cite{Savage:2008er}, CoGENT~\cite{Aalseth:2010vx} and very recent CDMS result\cite{Agnese:2013rvf} imply a light DM particle in the mass window $7-10~\GeV$ with spin-independent cross section $\sigma_{\rm SI} \sim  10^{-41}-10^{-40} {\rm\, cm}^{2}$. On the contrary, XENON100 and other experiments give the null
results. In order to check the possibility of a light DM, we also explore the corresponding mass window. However,
we found that the relic density constraint can not be satisfied in this region.

%
%

\section{Fine-tuning, unitarity and vacuum stability}

In this section, we analyze fine-tuning, unitarity and
vacuum stability problem. To preclude mixing
of S and the SM Higgs boson and the existence of cosmologically
problematic domain walls, we require $\langle S\rangle=0$~\cite{Gonderinger:2009jp}.
After the spontaneous symmetry breaking, the scalar potential has the following form,
\begin{eqnarray}
V(S,h)&=&\frac{M^2_{S}}{2}S^2+\frac{\lambda_{S}}{4}S^{4}+
\frac{\lambda_{hSS}}{8}h^2 S^{2}+\frac{\lambda_{hSS}}{4}hS^2 \nonumber\\
&+& \frac{\lambda}{4}h^4+\frac{m^{2}_{h}}{2}h^{2}+\lambda vh^3\;,
\end{eqnarray}
where the last two terms in above equation contribute to the tadpole
diagram, as is shown in the Fig.~\ref{fig:SVFT}. The vacuum stability of the model requires:
\begin{eqnarray}
 && \lambda>0, \quad\lambda_{S}>0,\nonumber\\
 &&  \rm{and}~\lambda^{2}_{hSS}<\lambda\lambda_{S}\quad\rm{for ~negative}\quad\lambda_{hSS}.
\end{eqnarray}

\subsection{Fine-tuning and unitarity}

With the two component dark matter fields entering into the SM, the $VC_{SM}$ in the right hand side
of Eq.~(\ref{eq:FTSM}) is replaced by $VC_{SV}$. The $O(N)$ symmetry preserved by the $S$ needs to be larger than $78$ for
the relic density permitted region $0.05<\lambda_{hSS,hVV}<0.35$. In this regime, the cutoff $\Lambda$
could be pushed to $10$ TeV, which greatly improves the fine-tuning problem.

We should mention that unlike the scalar DM case, the vector DM Lagrangian
is actually not renormalizable,
which may cause the unitarity problem when lack of a hidden sector which is responsible for its 'soft' mass $m_{V}$~\cite{Baek:2012se,Choi:2013eua}.
In order to restrict the vector DM Lagrangian as a reliable effective theory, we impose the partial wave unitarity bound in high energy limit $E\gg M_{V}$\cite{Lebedev:2011iq,Baek:2012se}
\footnote{Although the effective interaction is different from that of \cite{Lebedev:2011iq}, the unitarity bound has the same expression, since the discrepancy between them is
 eliminated by different definitions of the vector dark matter mass between the two papers.} :
\begin{eqnarray}\label{eq:unitarity}
E<\sqrt{\frac{16\pi}{\lambda_{hVV}}}\frac{M^{2}_{V}}{m_{V}}
\end{eqnarray}
Above this energy, additional contribution from hidden sector should be taken into account.
\begin{figure}[!htp]
  \centering
  \includegraphics[width=0.6\linewidth]{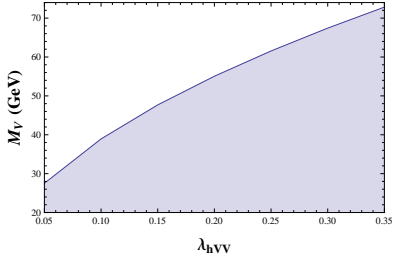}\\
  \caption{Exclusion regions of $M_{V}$ from the unitarity constraint.}\label{fig:MvU}
\end{figure}

We now analyze the reliable region for our model.
As is required by the relic abundance,
the Higgs-portal coupling $\lambda_{hVV}$ is of order O($10^{-1}$),
the cancellation between $m^{2}_{V}$ and $\frac{1}{4}\lambda_{hVV}v^2$
need to be very delicate in order to get small vector mass $M_{V}$.
Meanwhile, for light $M_V$, $E$ is required to be larger than $m_{h}$ in order for allowing the process $hh\rightarrow VV$,
 and for heavy $M_V$, $E$ is required to be larger than $2M_{V}$. In this situation,
 when the Eq.~(\ref{eq:unitarity}) is satisfied, the computation is valid in our model.
Based on the pervious analysis, to provide the right relic density, the Vector DM Higgs-portal coupling needs to live in the region $0.05\leq\lambda_{hVV}\leq0.35$.
The light $M_V$ region (shadow region in the Fig.~\ref{fig:MvU}.)
is precluded by the unitarity constraint.

\subsection{Stability}

\begin{figure}[!htbp]
\begin{center}
\includegraphics[width=0.49\linewidth]{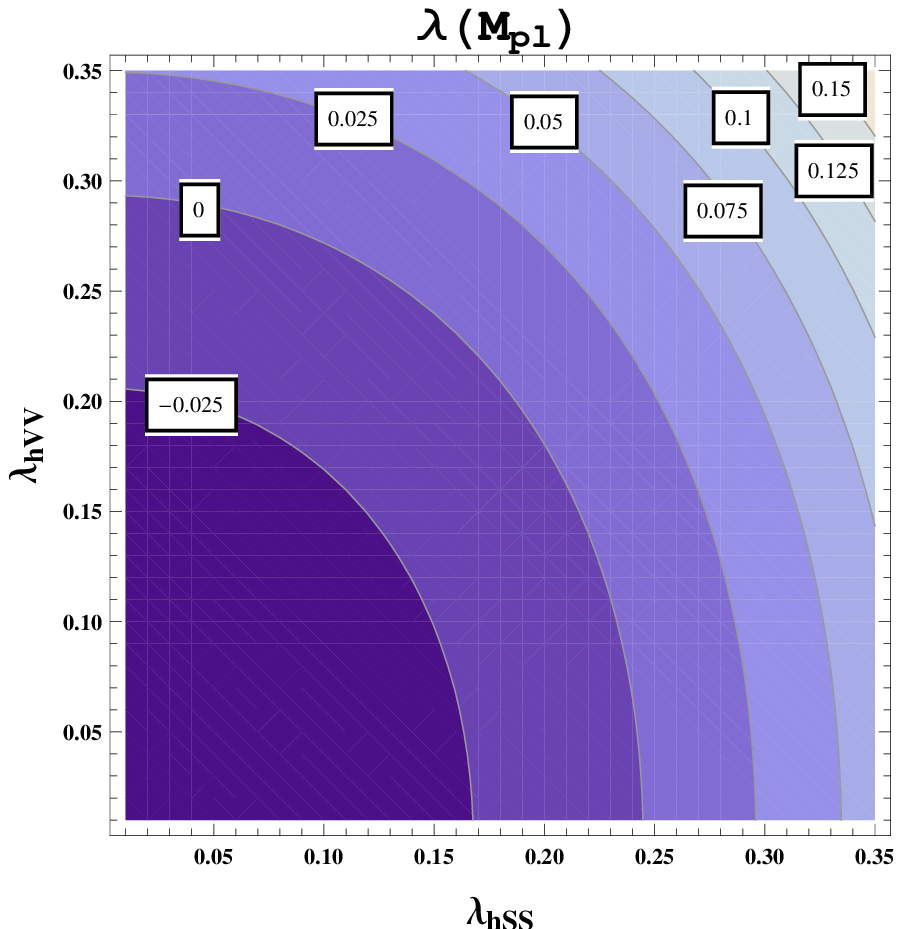}
\includegraphics[width=0.49\linewidth]{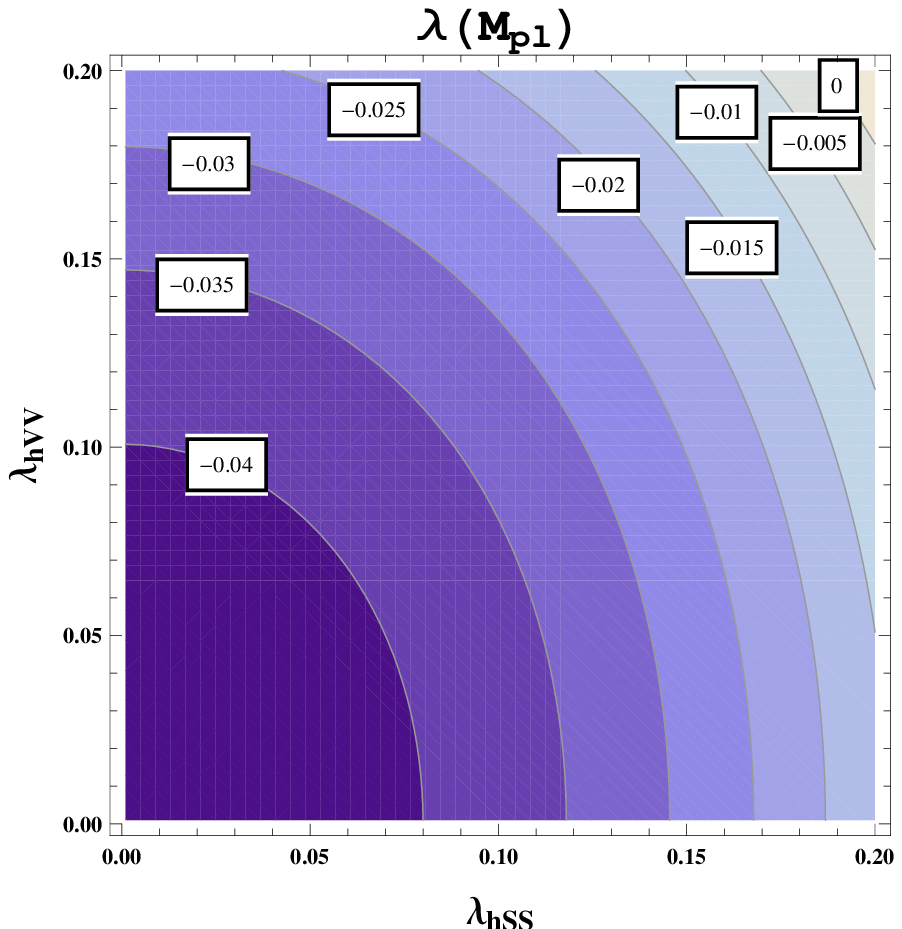}\\
\end{center}
\caption{Vacuum stability constraints in the $\lambda_{hSS}-\lambda_{hVV}$ plane for different parameter regions.}\label{fig:ST}
\end{figure}

Since both scalar and vector Higgs-portal coupling give positive contribution to the $\beta$ function of the Higgs quartic coupling, which will improve the
vacuum stability of the model. With one-loop $\beta$ functions list in Eq.~(\ref{eq:belam}), the value of the Higgs quartic coupling $\lambda(\mu)$
is solved up to Planck scale $M_{pl}\simeq 1.22\times10^{19}$ GeV. Fig.~\ref{fig:ST} displays the contour plot of $\lambda(\mu=M_{pl})$ in the $\lambda_{hSS}-\lambda_{hVV}$ plane. We take the regions of $\lambda_{hSS}$ and $\lambda_{hVV}$ which are consistent with the regions of relic density constraints. The vacuum stability constraint is denoted by the contour which is labeled by $\lambda(M_{pl})=0$, i.e., the regions above this contour are allowed~\cite{Degrassi:2012ry}. Furthermore, to preserve the perturbativity, the Higgs-portal couplings need to satisfy the requirement~\cite{Gonderinger:2009jp},
\begin{eqnarray}
&&\lambda_{hSS}\leq 8 \pi,\nonumber\\
&&\lambda_{hVV}\leq 8 \pi,\nonumber\\
&&\lambda_{S}\leq 2\pi/3 .
\end{eqnarray}
Which are permitted in the whole regions of parameter space which are shown in Fig.~\ref{fig:ST}.

\section{Conclusions}

In this work, we consider the possibility that dark matter candidates
could be composed of a real scalar $S$ and
a singlet vector $V$ (under the SM $U(1)$).
The two component dark matter model is obtained by adding them to the SM through the Higgs-portal.

To provide observed relic density measured by Planck, Higgs-portal couplings $\lambda_{hVV,hSS}$ need to live in the region $0.05<\lambda_{hVV,hSS}<0.35$.
If $\lambda_{hSS}=\lambda_{hVV}$, for the case $M_S>M_V$($M_S<M_V$), the region of $M_{V}$ permitted under the XENON100 bound satisfies the
unitarity constraint.
If $\lambda_{hVV}\neq\lambda_{hVV}$,
the suitable value of Higgs-portal couplings is $\lambda_{hVV}(\lambda_{hSS})=0.3(0.1)$ for $M_S>M_V$, which is almost precluded by unitarity constraint.
And the suitable values of Higgs-portal couplings $\lambda_{hVV}(\lambda_{hSS})$ is
0.05(0.35) for $M_V>M_S$, which is permitted by unitarity constraint for $M_V>$28 GeV,
where all our computations are reliable.
In addition, we would like to mention that invisible decay width constraint region~\cite{Djouadi:2011aa,Djouadi:2012zc} comes from LHC has already been excluded by XENON100,
see the right panel of Fig.~\ref{fig:reldlm1}(Fig.~\ref{fig:reldlm2}).
With the two component dark matter $S$ and $V$, the fine-tuning problem is relaxed,
and the stability of the Higgs potential is improved up to Planck scale in some
parameter regions with correct relic density.

\appendix
\section{One-loop beta functions}

The renormalization group equation and the $\beta$ functions are given by
\begin{eqnarray}\label{eq:belam}
\beta_{x}=\frac{dx}{d\log\mu}\;,
\end{eqnarray}
with
\begin{eqnarray}
&&\beta_{\lambda_{hVV}}=\frac{1}{16\pi^{2}}\frac{3}{2}\lambda^{2}_{hVV}\;,\\
&&\beta_{\lambda_{hSS}}=\frac{1}{16\pi^{2}}\lambda_{hSS}
(-\frac{9g^2_{2}+3g^2_{1}}{2}+6g^{2}_{t}+6\lambda\nonumber\\
&&\qquad\qquad+2\lambda_{hSS}+6\lambda_{S})\;,\\
&&\beta_{\lambda_{S}}=\frac{1}{16\pi^{2}}
(\frac{\lambda^2_{hSS}}{8}+18\lambda^{2}_{S})\;,\\
&&\beta_{\lambda}=\frac{1}{16\pi^{2}}\big(-\lambda(3g^2_{1}
+9g^2_{2}-12g^{2}_{t}-24\lambda)+\frac{3}{4}g^4_{2}\nonumber\\
&&\qquad+
\frac{3}{8}(g^2_{1}+g^2_{2})^2-6g^4_{t}
+\lambda^2_{hSS}+3\lambda^{2}_{hVV}
\big)\;,\\
&&\beta_{g_{2}}=\frac{1}{16\pi^{2}}(-\frac{19}{6}g^3_{2})\;,\\
&&\beta_{g_{1}}=\frac{1}{16\pi^{2}}(\frac{41}{6}g^3_{1})\;,\\
&&\beta_{g_{3}}=\frac{1}{16\pi^{2}}(-7g^3_{3})\;,\\
&&\beta_{g_{t}}=\frac{1}{16\pi^{2}}(\frac{9}{2}g^3_{t}-8g^2_{3}g_{t}-\frac{9}{4}g^{2}_{2}g_{t}
-\frac{17}{12}g^2_{1}g_{t})\;.
\end{eqnarray}

\end{document}